\documentclass[11pt]{article}
\usepackage{amsmath,amsthm,amsfonts,amssymb,mathrsfs,epsfig,bm}
\usepackage{pxfonts}
\usepackage{microtype}
\usepackage[usenames,dvipsnames]{xcolor}
\usepackage[colorlinks=true]{hyperref}
\usepackage{graphicx}
\usepackage[margin=1cm]{caption}

\usepackage{caption}
\usepackage[top=2cm, bottom=2cm, left=2cm, right=2cm, heightrounded,
marginparwidth=2.9cm, marginparsep=2mm]{geometry}
\parskip        \smallskipamount

\newtheorem{theorem}{Theorem}
\newtheorem{lemma}{Lemma}
\newtheorem{proposition}{Proposition}
\newtheorem{corollary}{Corollary}
\newtheorem{remark}{Remark}
\newtheorem{conjecture}{Conjecture}

\def\Z{\mathbb{Z}}

\def\R{\mathbb{R}}

\def\P{\mathbb{P}}
\def\E{\mathbb{E}}
\renewcommand{\phi}{\varphi}
\renewcommand{\epsilon}{\varepsilon}
\def\BB{\mathcal{B}}
\def\PP{\mathcal{P}}
\newcommand{\1}{\boldsymbol{1}}

\newcommand{\var}{\operatorname{var}}

\renewcommand{\Re}{\operatorname{Re}}

\def\no{\noindent}

\def\sto{<_{\text{st}}}
\def\AoI{\text{AoI}}    

\def\II{\mathcal I}
\def\JJ{\mathcal J}

\begin{document}

\title{\bf Age of information without service preemption}
\author{George Kesidis\thanks{gik2@psu.edu; Univ.\ of Pennsylvania, USA}
\and Takis Konstantopoulos\thanks{takiskonst@gmail.com; Univ.\ of Liverpool, UK; research supported by Cast.\ Co.\  IIS-75}
\and Michael A.\ Zazanis\thanks{zazanis@aueb.gr; Athens Univ.\ of Economics and 
Business, Greece}
}
\maketitle

\begin{abstract}
When designing a message transmission system, from the point of view of 
making sure that the information transmitted is as fresh as possible, 
two rules of thumb seem reasonable:
use small buffers and adopt  a last-in-first-out policy.
In this paper, we measure freshness of information using the 
``age of information'' performance measure.
Considering it as a stochastic process operating in a stationary regime,
we compute not just the first moment but the whole marginal distribution of
 the age of information (something important in applications) 
for  two well-performing systems. 
In neither case do we allow
for preemption of the message being processed because this may be
difficult to implement in practice.
We assume that the arrival process is Poisson and that the messages
have independent sizes (service times) with common distribution.
We use Palm and Markov-renewal theory to derive explicit results
for Laplace transforms. In particular, this approach can be used to analyze 
more complex last-in-first-out systems with larger buffer sizes.
\end{abstract}


\section{Introduction}

Traditionally, networked systems performance is measured with respect to
buffer sizes and delays. Relatively recently, there has been a shift in what is
considered important both in terms of design and performance. 
The introduction of the so-called ``age of information'' (usually abbreviated as
AoI), defined 
as the elapsed time since the information possessed by a monitor 
is generated and time stamped at the source, has received a lot of attention. 
The reason is simple: in several applications it is the freshness of information
that is important rather than the correct transmission of all packets. 
Examples include  virtual reality, online gaming,
weather reports, autonomous driving, stock market trading,
decision systems for an airplane, power systems, sensor/actuator systems and
other other ``cyber physical" systems.

We start by precisely defining the concept of AoI in general.
Consider a message processing  facility with one input stream
of arriving messages. The facility can be a single queue or a complex network system.
An arriving message has a certain positive ``size" (expressed
in time units and interpreted as processing or service time)
and three things can happen:
(i)  the message is immediately rejected upon arrival;
(ii) the message is accepted but rejected while in the system; 
(iii) the message is successfully transmitted  as soon as it is processed in its entirety. 
We are interested in the time that the latter will happen
in comparison to the time that the message arrives in the system.
If messages are labeled by integers in a way that the message with label $n \in \Z$ 
arrives at time $T_n \in \R$ and $T_m < T_n$ if $m< n$,
if $T_n+\Delta_n$ denotes the time at which message labeled $n$
leaves the system either by being rejected or successfully transmitted,
and if $\psi_n$ is a binary variable indicating the latter
($\psi_n=1$ if message $n$ is successful or $0$ if not), then we let
\begin{align}
D(t) &:= \sup\{T_n+\Delta_n:\, n \in \Z, T_n+\Delta_n \le t, \psi_n=1\},
\label{poi1}
\\
A(t) &:= \sup\{T_n:\, n \in \Z, T_n+\Delta_n \le t, \psi_n=1\},
\label{poi2}
\end{align}
and define the AoI at time $t$ by
\begin{equation}
\label{aoidef}
\alpha(t) := t-A(D(t)).
\end{equation}
Quite simply, $D(t)$ is the time of the last successful departure before $t$
and $A(D(t))$ is the arrival time of the message that departed at time $D(t)$.
This definition is quite general, that is, it does not depend on the 
details of the system design.

Typically, systems that adopt freshness of information as performance measure
should be designed so that its AoI ``be as small as possible''. 
The last phrase can mean several things. For example, it can mean that
the quantity $\alpha(t)$ is least for all $t$ under identical traffic
conditions. Or it could mean least in terms of an expectation or another
functional of the process.
Adopting AoI as a performance criterion immediately poses some simplifications over
traditional queueing theory performance criteria but also presents some new challenges. 

Suppose that the processing facility is a queue with one server and buffer consisting of
a finite or infinite number of cells.
One of the cells of the buffer is occupied by the message 
being processed 
and the rest of the cells are occupied by stored messages.
If we are free to design the buffer and the service policy as we wish,
how do we do this if we wish to keep the AoI ``as small as possible''?
 
It is reasonable to conjecture that every time a message arrives
we process it immediately (after all, we are not interested in obsolete information.)
That is, even if the server is busy at the moment
of arrival, the currently served message is immediately discarded and the 
new one starts being processed.  Systems working in this manner are service-preemptive.
It  also seems reasonable to serve messages in reverse order
of arrival: the most recent message must be served first (LIFO). 

One may thus conjecture that LIFO-preemptive (meaning: service preemptive) is 
``best''. 
But numerical examples and simulations show that this is false
depending on the model assumptions.
In particular, 
a single buffer system with
no service preemption (called $\BB_1$ below) has smaller AoI both
in expectation 
and stochastically under particular assumptions on the
message size distribution
\cite{Kosta17,Inoue19,KKZ19}.
In fact, we conjectured in \cite{KKZ19} that the so-called $\PP_2$ system
(see below and Figure \ref{p2fig} for the definition)
has lower AoI than $\BB_1$.
In this paper, {\em inter alia}, we resolve this issue.

The simplest systems with small-size buffer  
and no service preemption are defined
next.
One of them, denoted as $\BB_2$, is nothing else but a single-server queue with buffer of size $2$
and blocking. That is, an incoming message finding the buffer full is immediately discarded.
The other one, denoted as $\PP_2$, also works without service preemption.
An arriving message in $\PP_2$ finding the buffer full displaces or ``pushes out" the stored
message. See Figure \ref{p2fig} for a typical scenario in $\PP_2$.
\begin{center}
\includegraphics[width=5in]{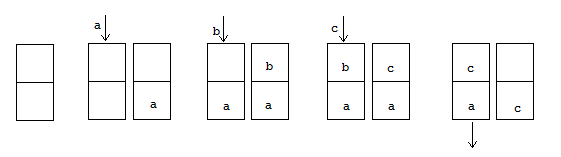}
\captionof{figure}{ The $\PP_2$ system. The  lower cell contains
the message being processed (if any).
Message $a$ arrives at an empty system and is immediately processed.
Message $b$ arrives and is stored in cell the upper cell. While the buffer still
contains $a$ and $b$, a third message $c$ arrives and immediately kicks $b$ out.
When $a$ completes being processed it departs and $c$ moves to the lower cell.}
\label{p2fig}
\end{center}
More generally, we define, for $n \ge 2$, $\BB_n$ and $\PP_n$ as follows.
The $\BB_n$ system is simply a single-server queue with buffer of size $n$,
operating under FIFO (First In First Out) policy and blocking of any incoming message
finding the buffer full.
The $\PP_n$ system, for $n \ge 2$, works as follows: 
messages are stored in an order that is reverse to their order of arrival;
so if there is a message being processed in cell 1 at time $t$, 
the message in cell 2 arrived last before $t$ while the message in cell $n$
is the oldest; a new message arriving at a full buffer is always stored in cell 2, displacing 
the other messages upwards and pushing out the one sitting in cell $n$ (oldest one).
For $n \ge 2$, $\PP_n$ has no service preemption.

Note that in some cases, queuing and service
policies may be subject  to
technological constraints (e.g., whether non-FIFO dequeueing or preemptive
service policies are feasible, and what aspects of the queueing
system are observable).
In particular,
$\BB_2$ may be the best policy
if LIFO dequeueing, queue pushout, and service preemption
are all infeasible but blocking is, cf. Section \ref{secbest}.

For $n=1$, the system $\BB_1$ is simply a single-buffer blocking queue:
an arriving message is immediately rejected if there is a message in the system.
The $\PP_1$ system is the single-buffer push-out system:
an arriving message pushes out the currently processed message and
takes its place.
The $\PP_1$ and $\BB_1$ systems were analyzed under very general conditions in \cite{Inoue19,KKZ19}.

Of all systems $\PP_n$, $\BB_n$, $n \ge 1$, the $\PP_1$ system is special because it 
is the only one for which service preemption is allowed.

In this paper, we analyze the AoI processes 
and derive the stationary AoI distribution for 
the $\BB_2$ (in Section \ref{b2sec}) and $\PP_2$ (in Section \ref{p2sec})  systems
under a Poisson arrival process but generally distributed message sizes.
The distribution for $\PP_2$ case was previously derived in 
\cite{Inoue19}.
The technique used herein is different from that of \cite{Inoue19,KKZ19} 
and is easily generalizable to other systems,
particularly to LIFO ``dequeuing" policies $\PP_n$ for $n>2$ \cite{P3}.
Indeed, we use the classical embedding technique, valid for queueing systems
with Poisson arrivals (see, e.g.\ \cite{CIN}), giving that the system
sampled at certain epochs (the departures of successful messages in
our case) has a Markovian property.
This, of course, depends crucially on the Poisson property of the arrival process,
so, for more general arrival processes (as in the case of \cite{Inoue19,KKZ19})
we need a different method that is substantially more complex and shall
not be considered herein.

Under our assumptions, and because we consider finite buffers,
it holds that there is a unique stationary version of the stochastic process
$\alpha$ in all systems considered. We will not offer any reasons for this technical result, but only 
point out that even existence may not hold if the arrival and message size processes
are neither independent nor renewal, and point out the difficulties by 
referring to \cite{BB}.
We shall always be considering the stationary version.
Hence $\alpha(t)$ has the same distribution for all $t$.
This is precisely what we are interested in describing.
We note that computing the expectation is, in general, not that much
easier than deriving the whole distribution. We also note that deriving the
distribution is essential in case that we are interested not just in maintaining
a low AoI on the average but also in maintaining the tail of the probability 
distribution small. 

While our techniques are generalizable to all $\PP_n$ and $\BB_n$, for any positive
integer $n$, we only work with $n=2$ for the following reason.
We claim that the  random variables 
$\alpha_{\PP_n}(t)$, $\alpha_{\BB_n}(t)$, $n \ge 3$, are
stochastically larger than 
$\alpha_{\PP_n}(t)$, $\alpha_{\BB_n}(t)$, $n \in \{1,2\}$.
Although, currently, we have no full proof of this fact, we have enough evidence to
pose this as a conjecture. We partially prove and justify this conjecture
in Section \ref{secbest}.

Throughout the paper, we let $\lambda$ be the rate of the (Poisson) arrival
process and $G$ the distribution of a typical message size $\sigma$,
a random variable that is positive with probability $1$ and has finite expectation
denoted by $1/\mu$. We thus only assume that $\lambda >0$ and $\mu>0$
(but $\sigma$ may have infinite variance).
It is assumed that the message sizes are i.i.d.\ copies of $\sigma$ and independent
of the arrival process. The ratio $\rho=\lambda/\mu$ is referred to as traffic intensity.

The main results of the paper are Theorems \ref{thmB2} and \ref{thmP2}.
We here present a special case.
When the message sizes are i.i.d.\ exponential, having
(for notational convenience) rate $\mu=1$,
we shall show {as a corollary} that, in steady-state,
the value of $\AoI$ at some (and hence any) point of
time, has density
\begin{align}
f_{\BB_2}(t) &=c\,(q(t) e^{-t} + e^{-\lambda t}) ,   \label{fb}
\\
f_{\PP_2}(t) &= q_1(t) e^{-t} + q_2(t) e^{-(\lambda+1)t} - \frac{\lambda}{\lambda-1} e^{-\lambda t},   \label{fp}
\end{align}
in the $\PP_2$ and $\BB_2$ cases, respectively, where
\begin{align*}
c=  \frac{\lambda}{(\lambda^2+\lambda+1)(\lambda-1)^2}, & \quad 
q(t) =\tfrac12 \lambda(\lambda-1)^2 t^2+\lambda(\lambda-1)t-1,\quad 
\\
q_1(t) = \frac{( {\lambda}^{3}+{\lambda}^{2}-2\,\lambda ) t-{\lambda}^{2}+\lambda+3}{(\lambda^2+\lambda+1)(\lambda-1)}, &
\quad 
q_2(t) = \frac{( {\lambda}^{2}+\lambda ) t+{\lambda}^{2}+3\,\lambda+3}{\lambda^2+\lambda+1},
\end{align*}
when $\lambda \neq 1$; while, for $\lambda=1$, the densities become
\[
f_{\BB_2}(t) = \tfrac13(t^2+t)e^{-t},\quad
f_{\PP_2}(t) =   \tfrac13 ( 7+2t) e^{-2t} + \tfrac13  (6t-7) e^{-t}.
\]
(For general $\mu$, simply replace $\lambda$ by $\lambda/\mu$ and $t$
by $\mu t$ in the foregoing expressions.)
These results are by themselves generalizations of what is already in the literature.
For instance, in \cite{Kosta17} the expectations for $\BB_2$ and $\PP_2$,
see \eqref{meanexpp2} and \eqref{meanexpb2} in Corollaries \ref{corob2exp} 
and \ref{corop2exp} {\em infra},
have been computed under the same probabilistic assumptions.
Notice that, in particular, 
\[
\lim_{\lambda \to \infty} f_{\BB_2}(t) = \frac12 t^2 e^{-t},
\quad
\lim_{\lambda \to \infty} f_{\PP_2}(t)  = t e^{-t},
\]
and the limiting functions are probability densities as well.
The first one is Erlang(3) and the second Erlang(2).
Similar limits can be obtained for $\PP_n$ and $\BB_n$, without computations,
simply by considering the system dynamics.
In fact, we can interpret $\BB_\infty$ as a single server queue with infinite buffer operating under
the FIFO discipline. From the point of view of AoI, this system is worst and should not
be considered. Our practical rule of thumb says that one should store
at most one message and discard everything beyond that, insofar as keeping AoI low is
the goal. This rule of thumb may not be always optimal but it is frequently close to optimal.

The rest of the paper is organized as follows.
We explain the basis of the technique used in Section \ref{basis}
and see why it is absolutely general, as long as the system, possibly a network, 
has Poisson 
arrivals.  The $\BB_2$ system is considered in Section \ref{b2sec}
and the $\PP_2$ in Section \ref{p2sec}.
Section \ref{compa} contains a number of interesting observations.
First it provides comparisons of the systems analyzed in this paper,
together with the systems $\PP_1$ and $\BB_1$ (analyzed under more general
assumptions in \cite{Inoue19,KKZ19}). Second, it explains what the limits
of the AoI are when $\lambda \to \infty$.
Third, it gives a way to understand the AoI for $\PP_1$ when
the message size is deterministic.
In this case the Laplace transform is not invertible (we only invert it
numerically) but the moments
have an interesting combinatorial explanation.
Fourth, we justify our observation that the ``best system''
is among $\PP_n, \BB_n$, for $n=1,2$.
We conclude with some words for future work in
Section \ref{sec:summary}.

\section{Basic framework}\label{basis}
We discuss the technique used in the analysis for all systems described
in the introduction from the point of view of the distribution of the 
age of information.
By this phrase, we will always mean that the age of information process
$\alpha(t)$, $t \in \R$, is stationary and that we shall be interested in
the distribution of $\alpha(t)$ for some, and hence all, $t$ which will be taken 
to be the point $t=0$.
The goal is to derive a ``fixed point equation'' for $\alpha(0)$,
see equation \eqref{stat-palm-aoi} below.
We note right away, that the present analysis is different than that of paper 
\cite{KKZ19,Inoue19}
 as we take advantage of an embedded Markov chain.
The arrival process is always taken to be Poisson process on $\R$ (=time)
with rate $\lambda$. 
As mentioned above, $T_n$ is the arrival time of message
labeled $n$. Its size is $\sigma_n$. The collection of message sizes
are i.i.d.\ and independent of the arrival process. Let 
\[
G(x)=\mathbb{P}(\sigma_1 \leq x)
\] 
be the distribution function of the typical size
and let $1/\mu$ be its expectation, assumed to be finite. 
Also assume that $G(0)=0$.
Abusing notation, we shall let $G$ denote the probability measure defined by the
function $G(x)$ and by $\hat G(s)$ its Laplace transform: 
\[
\hat G(s)
= \int_0^\infty e^{-sx} dG(x).
\]
Recall that $T_n+\Delta_n$ is defined  as the time at which message $n$ departs
either because it was pushed out or rejected or because it  was successfully
processed ($\psi_n=1$ in the latter case). 
See discussion around \eqref{poi1} and \eqref{poi2} where these symbols 
were introduced.
Then the number of messages in the system at time $t \in \R$ is given by
\[
\xi(t)= \sum_{n \in \Z} \1_{T_n \le t < T_n+\Delta_n}.
\]
Note that if the message is immediately rejected then $\Delta_n=0$ and so this
message does not contribute to $\xi$.
We let 
\[
\{S_m, m \in \Z\} := \{T_n+\Delta_n:\, n \in \Z,\, \psi_n=1\},
\]
and, thinking of the two sets as sequences, 
$\{S_m\}$ is a subsequence of $\{T_n+\Delta_n\}$
and is enumerated so that $S_{m_1} < S_{m_2}$ if $m_1 < m_2$.
We note that $\xi$ is right-continuous for all $t$.
Recalling the notions of Markov renewal and semi-Markov processes,
see, e.g., \cite[VII.4]{ASM},
our first observation is:
\begin{lemma}
\label{semim}
For both $\BB_2$ and $\PP_2$ cases, the process 
$\xi(t)$, $t \in \R$, is a semi-Markov process \cite[Ch.10]{CIN} 
with respect to the points $S_n$, $n \in \Z$. 
Moreover, the distribution of $\xi$ is the same
in both $\BB_2$ and $\PP_2$ cases.
\end{lemma}
This follows easily by standard arguments in queueing theory,
for instance in the analysis of a queue with Poisson arrivals;
see, e.g., \cite[Ch.\ 6, Sec.\ 5]{CIN}.
Thus, $\xi$ does not ``see'' the difference between $\BB_2$ and $\PP_2$.
The distinction between these two will become important in the next section when we discuss the details about $\alpha$ in each case.

We further assume that  the arrival process together with the process $\xi$ are
stationary under a probability measure $\P$.
(This assumption is non-vacuous; we shall not elaborate on this further but
refer the reader to \cite{BB} for an exposition of techniques used to establish it.)

\begin{figure}[h]
\begin{center}
     \includegraphics[width=0.95\textwidth]{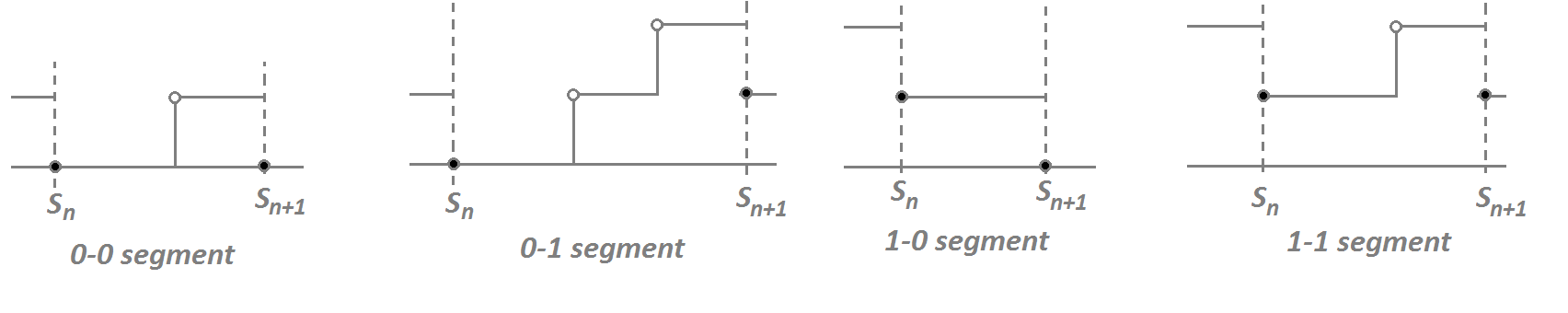}
  \caption{ What $\xi(t)$ looks like when $S_n \le t < S_{n+1}$,
regardless of the policy used.}
    \label{fig:segments}
\end{center}
\end{figure}

We refer to the intervals $[S_n, S_{n+1})$ as {\em segments}
and split the paths of $\xi$ into union of paths over segments.
See Figure \ref{fig:segments}.
By convention, we assume that the segment labelled $0$ contains the point $t=0$.
Denote by $\mathbb{P}^0$ the Palm probability of $\P$ 
with respect to the point process $\{S_n\}$. We refer to  \cite{BB}  for this concept. Intuitively,
$\P^0$ is $\P$ conditional on the event that $0 \in \{S_n, n \in \Z\}$.
Hence $\P^0(S_0=0)=1$.
Let
\[
K_n := \xi(S_n), \quad n \in \Z.
\]
The sequence $\{K_n\}$ is a Markov chain with state space $\{0,1\}$ 
while $\{(S_n,K_n)\}$ is the Markov renewal
sequence  \cite{ASM, CIN} associated to the semi-Markov process $\xi$.
The latter has transition kernel
\[
Q_{ij}(x):= \mathbb{P}^0(S_{n+1}-S_n \leq x, K_{n+1}=j \, | \, K_n=i),
\quad i, j \in \{0,1\},
\]
explicitly given by
\begin{equation} \label{SMKernel}
\left[ \begin{array}{cc} Q_{00}(x) & Q_{01}(x) \\ Q_{10}(x) & Q_{11}(x) \end{array} \right] \;=\;
\left[ \begin{array}{cc}
\int_0^x \left( 1-e^{-\lambda (x-u)}\right)e^{-\lambda u}  dG(u) 
& \int_0^x \left(1-e^{-\lambda(x-u)}\right)\left(1-e^{-\lambda u}\right) dG(u) \\ & \\
\int_0^x e^{-\lambda u} dG(u) & \int_0^x \left(1-e^{-\lambda u}\right) dG(u) \end{array} \right],
\end{equation}
as follows easily by considering the cases of Figure \ref{fig:segments}.
Letting $x\rightarrow \infty$ in \eqref{SMKernel} we obtain the transition matrix for the Markov chain $\{K_n\}$,
\[
\left[ \begin{array}{cc} Q_{00}(\infty) & Q_{01}(\infty) \\ Q_{10}(\infty) & Q_{11}(\infty) \end{array} \right] \;=\;
\left[ \begin{array}{cc} \hat{G}(\lambda) & 1-\hat{G}(\lambda) \\\hat{G}(\lambda) & 1-\hat{G}(\lambda)  \end{array} \right],
\]
from which it is evident that $K_n, n \ge 1$, is not just Markovian but also
a sequence of independent Bernoulli random variables with
\begin{equation} \label{PK01}
\mathbb{P}^0(K_n=0) =  \hat{G}(\lambda)=1-\mathbb{P}^0(K_n=1).
\end{equation}
Figure \ref{fig:segments} shows the four different types of segments depending on the values of $K_n$ and $K_{n+1}$.
We next  define 
\[
\Phi_i(s):=\mathbb{E}^0[e^{-s (S_1-S_0)}|  K_0=i],
\] 
and, using the kernel (\ref{SMKernel}), we obtain
\begin{align}
\Phi_0(s) &=\int_0^\infty e^{-s x} dQ_{00}(x) + \int_0^\infty e^{-s x} dQ_{01}(x)
= \frac{\lambda}{\lambda+s} \,\hat{G}(s),   \label{Phi0} 
\\ 
\Phi_1(s) &= \int_0^\infty e^{-s x} dQ_{10}(x) + \int_0^\infty e^{-s x} dQ_{11}(x) 
= \hat{G}(s).    \label{Phi1}
\end{align}
From (\ref{Phi0}), (\ref{Phi1}), and (\ref{PK01}) we obtain the Laplace transform
of the segment length:
\begin{equation*}   
\Phi(s) \; := \; \mathbb{E}^0 [ e^{-s(S_1-S_0)} ] \;=\;  \left( 1-\hat{G}(\lambda) + \hat{G}(\lambda) \frac{\lambda}{\lambda+s} \right) \, \hat{G}(s).
\end{equation*}
From this, we obtain the mean length of a segment as 
\begin{equation} \label{mean_length_seg}
\mathbb{E}^0[S_{1}-S_0] \;=\; \frac{1}{\mu} + \frac{\hat{G}(\lambda)}{\lambda}.
\end{equation}

We shall henceforth use the abbreviation $\E(X; A)$ for the expectation of
a random variable $X$ on the event $A$, that is, the quantity $\E(X \1_A)$.
The following result depends entirely on the semi-Markov property of $\xi$.
\begin{proposition}
\label{prop1}
The random variable $\alpha(0)$ satisfies
\begin{equation} \label{stat-palm-aoi}
\mathbb{E}[e^{-s\alpha(0)}] \;=\; \frac{\lambda}{s} \cdot \frac{ \mathbb{E}^0 [ e^{-s\alpha(0)} ;{K_0=0}]\,
\left( 1- \frac{\lambda}{\lambda+s}\hat{G}(s)\right) +  \mathbb{E}^0 [ e^{-s\alpha(0)} ;{K_0=1}]\, \left( 1- \hat{G}(s) \right) }{ \frac{\lambda}{\mu} + \hat{G}(\lambda)}.
\end{equation}
\end{proposition}
\begin{proof}
The Palm inversion formula \cite{ASM, BB} applied to the $\P$-stationary process
$\alpha$ gives
\begin{equation} \label{palminv}
\mathbb{E}[e^{-s\alpha(0)}] = \frac{\mathbb{E}^0[\int_{S_0}^{S_{1}} e^{-s \alpha(t)} dt ]}{\mathbb{E}^0[S_{1}-S_0]}.
\end{equation}
Take a look at  \eqref{aoidef} and notice that the process $\alpha$ is right-continuous.
Its set of discontinuities is $\{S_n\}$.
Moreover, it increases at unit rate on each segment:
\begin{equation}			\label{aaa}
\alpha(t) = \alpha(S_n)+t-S_n, \hspace{0.1in} \mbox{ for $t\in [S_n,S_{n+1})$.}
\end{equation}
To see this, notice that, for $S_n \le t < S_{n+1}$, we have
$D(t) = D(S_n) =S_n$, by the definition of $D$ in \eqref{poi1},
and so $A(D(t))=A(D(S_n)) = A(S_n)$.
Since, from the definition \eqref{aoidef},  $\alpha(t) = t-A(D(t))$ for all $t$,
we have
\[
\alpha(t) =  t-A(S_n),
\]
whenever $S_n \le t < S_{n+1}$.
Writing this for $t=S_n$, we have 
\[
\alpha(S_n) = S_n - A(S_n),
\]
and so \eqref{aaa} is obtained by subtracting the last two displays.
In particular, $S_0=0$ and
$\alpha(t) = \alpha(0) +t$ for $t \in [S_0,S_1)$, $\P^0$-a.s. Hence, for $i=0,1$,
\begin{equation} \label{age_integral}
\mathbb{E}^0 \left[ \int_{S_0}^{S_{1}} e^{-s \alpha(t)} dt  ; {K_0=i} \right] =
\mathbb{E}^0 \left[e^{-s \alpha(0)} \, \frac{ 1- e^{-s S_{1}} }{s} ; {K_0=i} \right]
= 
s^{-1} \Phi_i(s)
\mathbb{E}^0[e^{-s \alpha(0)}  ; {K_0=i} ],
\end{equation}
where the last equality follows from the fact that
$\alpha(0)$ and $S_{1}-S_0$ are conditionally independent given $\{K_0=i\}$, 
a consequence of the semi-Markov structure of the process $\{\xi(t)\}$, see Lemma
\ref{semim}.
Using expressions (\ref{Phi0}) and (\ref{Phi1}) and adding the terms in
\eqref{age_integral} we obtain the  numerator of \eqref{palminv}. 
The denominator is given by
\eqref{mean_length_seg}.
This shows the validity of (\ref{stat-palm-aoi}).
\end{proof}

\begin{remark}\label{rem1} \rm
It should be clear that \eqref{stat-palm-aoi} holds for a much larger class of systems
with one (or several independent) Poisson arrival process(es).
For example, we may define $\BB_n$ to be an extension of $\BB_2$ when 
the buffer has $n$ cells where messages are stored according to the order of their
arrivals and a message arriving to a full buffer is immediately rejected (the so-called
M/GI/1/$n$ queue). On the other hand, we may define $\PP_n$ to be an extension
of $\PP_2$: messages are stored in an order that is reverse to their order of arrival;
so if there is a message being processed in cell 1 at time $t$, 
the message in cell 2 arrived last before $t$ while the message in cell $n$
is the oldest; a new message arriving at a full buffer is always stored in cell 2, displacing 
the other messages upwards and expels the one sitting in cell 1 (oldest one).
In both $\BB_n$ and $\PP_n$, the process $\xi$ is semi-Markov and thus 
Proposition \ref{prop1}, depending only on this semi-Markov property,
applies and formula \eqref{stat-palm-aoi} holds.
In fact, one can assert that proposition holds for networks
with i.i.d.\ message sojourn times, e.g., due to a single bottleneck server.. 
We shall not attempt to formalize this further in this paper.
\end{remark}

\section{The $\BB_2$ system} \label{b2sec}
Recall that the $\BB_2$ system is the same as a single server queue with buffer
size 2. Under our Poisson assumption for the arrival process and i.i.d.\ assumptions 
for message sizes, this is further denoted by M/GI/1/2 in standard queueing theory.
We are, however, interested not in the number of messages in the system
neither on message delays but, rather, on the age of information process
$\alpha$. Assuming that $\alpha$ is stationary,
we compute the Laplace transform of $\alpha(0)$ under $\P$ by using
\eqref{stat-palm-aoi}
which requires knowledge of 
$\E^0[e^{-s\alpha(0)} ; {K_0=j}]$, $j=0,1$.
To obtain the latter,
we consider the segment $[S_{-1}, S_0)$ and  further condition on $K_{-1}$
and summarize the results in Lemma \ref{lemma:block} below.
In what follows, we let $\tau, \sigma$ be two independent
random variables, where $\tau$ is exponential with rate $\lambda$ and $\sigma$
has distribution $G$. 
\begin{lemma} \label{lemma:block}
For $\BB_2$, 
\begin{align} 
\mathbb{E}^0 [e^{-s \alpha(S_0)} ;{K_{-1}=0,K_0=0}] 
&= \hat{G}(\lambda)\hat{G}(s+\lambda) ,					\label{aoi00}
\\ 
\mathbb{E}^0[e^{-s \alpha(S_0)} ;{K_{-1}=0,K_0=1}] 
&= \hat{G}(\lambda) \left(\hat{G}(s) - \hat{G}(s+\lambda)\right) 	,				\label{aoi01}  
\\ 
\mathbb{E}^0[e^{-s \alpha(S_0)} ;{K_{-1}=1,K_0=0}] 
&= \frac{\lambda}{\lambda-s}\,\left(\hat{G}(s)-\hat{G}(\lambda)\right)  \hat{G}(s+\lambda), 	\label{aoi10}
\\ 
\mathbb{E}^0[e^{-s \alpha(S_0)} ;{K_{-1}=1,K_0=1}] 
&= \frac{\lambda}{\lambda-s}\,\left(\hat{G}(s)-\hat{G}(\lambda)\right)  \left(\hat{G}(s) - \hat{G}(s+\lambda)\right).			\label{aoi11}
\end{align}
\end{lemma}
\begin{proof}
Recall that the $K_n$ are i.i.d.\ with distribution \eqref{PK01}: 
$\P^0(K_n=0)=\hat G(\lambda)$. We shall consider the four cases separately
and, in each case, we shall be referring to the definition \eqref{aoidef}
to figure out what $\alpha(0)$ is.

\no
{\em Case 1.} $K_{-1}=0,K_0=0$.
Observe $\alpha(0) = S_0-T_0$, see Figure \ref{fig:lm1}. But
\[
\mathbb{P}^0(S_0-T_0 \in dx \mid K_0=0,K_{-1}=0) \;=\; \mathbb{P}(\sigma \in dx \mid \sigma < \tau ) \; = \; \frac{e^{-\lambda x} dG(x)}{\hat{G}(\lambda)},
\]
and so
\begin{eqnarray*}
\mathbb{E}^0[e^{-s \alpha(S_0)} ;{K_{-1}=0,K_0=0}] 
&=& \int_0^\infty e^{-sx} \frac{e^{-\lambda x}}{\hat{G}(\lambda)} dG(x) \left(\hat{G}(\lambda)\right)^2 \;=\; \hat{G}(\lambda) \hat{G}(s+\lambda).
\end{eqnarray*}
\begin{center}
\begin{figure}[h]
  \begin{center}
    \includegraphics[width=4.0in]{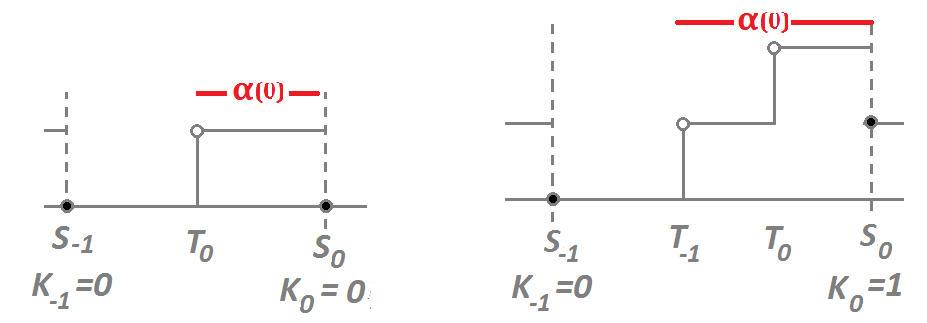}
  \end{center}
  \caption{ The segment $[S_{-1},S_0)$ when $K_{-1}=K_0=0$ (left) and $K_{-1}=0$, $K_0=1$ (right), with $S_0=0$.
}
  \label{fig:lm1}
\end{figure}
\end{center}

\no
{\em Case 2.} $K_{-1}=0,K_0=1$.
We have $\alpha(0) = S_0-T_{-1}$, see Figure \ref{fig:lm1}. Since
\[
\mathbb{P}^0(S_0-T_{-1}\in dx\mid K_0=1 , K_{-1}=0) \;=\; \mathbb{P}(\sigma \in dx \mid \tau<\sigma) \;=\; \frac{\left(1-e^{-\lambda x}\right) dG(x)}{1-\hat{G}(\lambda)}
\]
we obtain
\begin{eqnarray*}
\mathbb{E}^0[e^{-s \alpha(S_0)} ;{K_{-1}=0,K_0=1}] &=& 
\mathbb{P}^{0}(K_{-1}=0,K_0=1) 
\mathbb{E}^0[e^{-s \alpha(S_0)}\mid K_{-1}=0,K_0=1] 
\\
&& \hspace{-1.4in} =\; 
\hat{G}(\lambda) 
\int_0^\infty e^{-sx} \left(1-e^{-\lambda x}\right)dG(x)  
\;=\;
\hat{G}(\lambda) \left(\hat{G}(s)-\hat{G}(s+\lambda)\right).
\end{eqnarray*}

\no
{\em Case 3.} $K_{-1}=1,K_0=0$. To figure out $\alpha(0)$ we are
here forced to consider two consecutive segments. We then have
\[
\alpha(0) = \left(S_0-S_{-1}\right) + \left( S_{-1}-T_0\right),
\] 
see Figure \ref{fig:lm2}.  
Note that
\[
\text{ $S_0-S_{-1}$ and $S_{-1}-T_0$ are conditionally independent  given $K_{-1}=1$}
\]
with 
$\mathbb{P}^0(S_0 - S_{-1}\in dx; K_0=0 \mid K_{-1}=1) = \mathbb{P}(\sigma \in dx; \sigma < \tau)$
and  
$\mathbb{P}^0(S_{-1}-T_0\in dx; K_{-1}=1 \mid K_{-2}=0) = \mathbb{P}(\sigma - \tau \in dx; \sigma > \tau)$, respectively. Thus
\begin{eqnarray*}
\mathbb{E}^0[e^{-s\alpha(0)} ;{K_{-1}=1,K_0=0}] &=& \mathbb{E}[e^{-s\sigma};{\sigma < \tau}] \, \mathbb{E}[e^{-s(\sigma - \tau)};{\sigma > \tau}] \\
&& \hspace{-1in} \;=\; \mathbb{E}[e^{-s\sigma} e^{-\lambda \sigma}] \,  \mathbb{E}[e^{-s\sigma} \int_0^\sigma \lambda e^{-(\lambda-s) t}dt] \;=\;
\hat{G}(s+\lambda)  \frac{\lambda}{\lambda -s } \left(\hat{G}(s) - \hat{G}(\lambda) \right).
\end{eqnarray*}
\begin{center}
\begin{figure}[h]
  \begin{center}
    \includegraphics[width=5.0in]{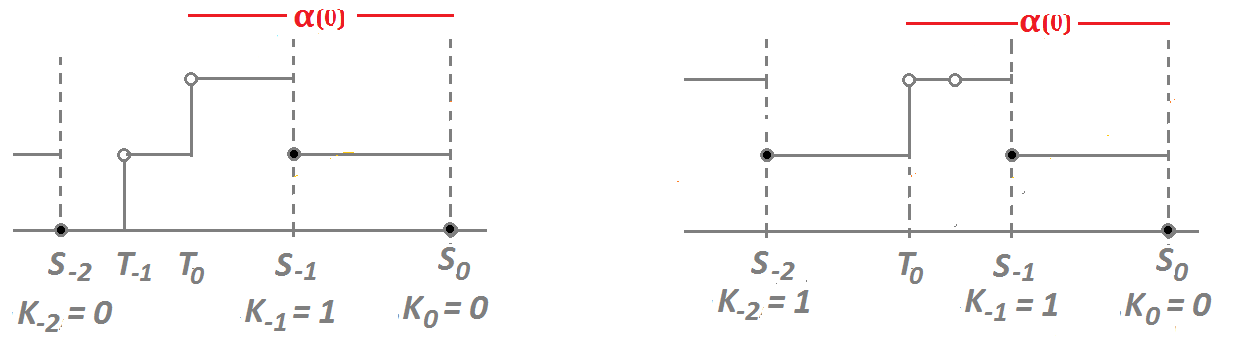}
  \end{center}
  \caption{
The segments $[S_{-2},S_{-1})$, $[S_{-1},S_0)$  when $K_{-1}=1,K_0=0$ in the case where $K_{-2}=0$ (left) and $K_{-2}=1$ (right).}
  \label{fig:lm2}
\end{figure}
\end{center}
\no
{\em Case 4.} Again, we have to consider two consecutive segments to realize that
\[
\alpha(0)= S_0-T_{-1} = \left(S_{-1}-T_{-1}\right) + \left( S_0-S_{-1} \right),
\]
see Figure \ref{fig:lm25}.
The two random variables 
\[
\text{$\left(S_{-1}-T_{-1}\right)$
and $\left( S_0-S_{-1}\right)$ are conditionally independent given that $K_{-1}=1$}
\] 
and thus
\begin{center}
\begin{figure}[h]
  \begin{center}
    \includegraphics[width=6.0in]{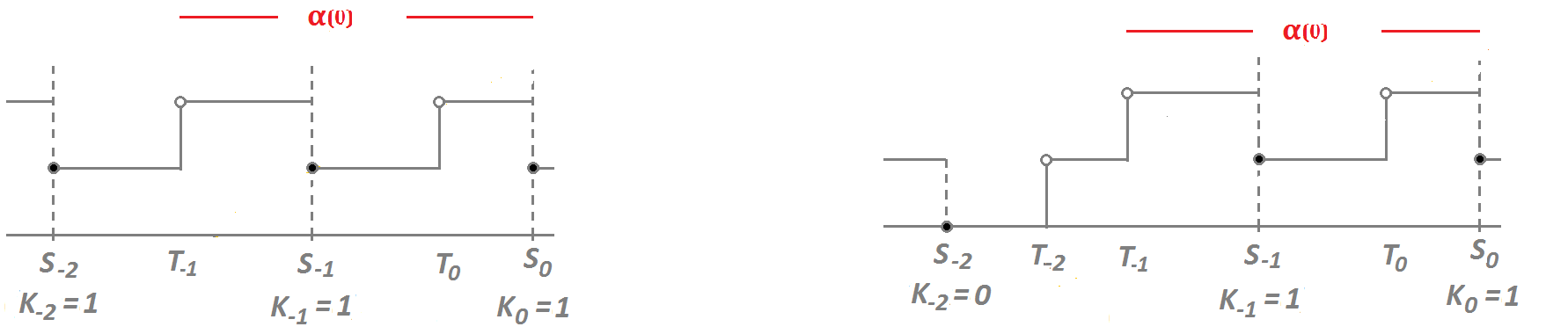}
  \end{center}
  \caption{ The segments $[S_{-2},S_{-1})$, $[S_{-1},S_0)$  when $K_{-1}=1,K_0=1$ in the case where $K_{-2}=0$ (left) and $K_{-2}=1$ (right).}
  \label{fig:lm25}
\end{figure}
\end{center}
\begin{eqnarray*}
\mathbb{E}^0[e^{-s\alpha(0)};{K_{-1}=1,K_0=1}] &=&
\mathbb{E}^0[e^{-s\left(S_{-1}-T_{-1}\right) - s \left( S_0-S_{-1} \right)};{K_{-1}=1,K_0=1}] \\
&& \hspace{-1.4in} =\; \mathbb{E}[e^{-s(\sigma-\tau)} ;{\sigma > \tau}] \,  \mathbb{E}[e^{-s \sigma} ;{\sigma > \tau}] \;=\;
\mathbb{E} \left[ \int_0^\sigma e^{-s(\sigma-t)} \lambda e^{-\lambda t} dt \right] \, \mathbb{E} \left[ e^{-s \sigma} \left( 1- e^{-\lambda \sigma} \right)\right]  \\
&& \hspace{-1.4in} =\; \frac{\lambda}{\lambda - s} \left(\hat{G}(s) - \hat{G}(\lambda) \right) \, \left( \hat{G}(s) - \hat{G}(s+\lambda) \right).
\end{eqnarray*}
This completes the proof.
\end{proof}

Define
\begin{equation}
\label{GI}
\hat{G}_I(s) =  \frac{1-\hat{G}(s)}{s}\mu.
\end{equation}
This is the Laplace transform of a probability measure $G_I$ that
is well-known in renewal theory: 
If we consider a renewal process with points, say, $Z_n$, $n \in \Z$,
such that $Z_0=0$ and $Z_{n+1}-Z_n$ having distribution $G$,
then there it has a stationary version (with no point at $0$)
and in such a way that $Z_1$ has distribution $G_I$.


\begin{theorem}\label{thmB2}
 For $\BB_2$, 
the Laplace transform of the stationary Age of Information is given by
\begin{equation} \label{Fifo-decomposition}
\mathbb{E}[e^{-s\alpha(0)}] \;=\;\hat{G}(s) \,  \left( \hat{G}(\lambda) + \lambda \frac{\hat{G}(s)- \hat{G}(\lambda)}{\lambda-s} \right) \,
\left( \frac{\hat{G}(\lambda)}{\frac{\lambda}{\mu}+\hat{G}(\lambda)} \frac{\lambda}{\lambda +s} \frac{\hat{G}(s+\lambda)}{\hat{G}(\lambda)} +  \frac{\frac{\lambda}{\mu}}{\frac{\lambda}{\mu}+\hat{G}(\lambda)} \hat{G}_I(s) \right).
\end{equation}
\end{theorem}

\begin{proof}
Summing  \eqref{aoi00} and \eqref{aoi01} we obtain
\begin{eqnarray}  \nonumber
\mathbb{E}^0[e^{-s \alpha(S_0)} ;{K_0=0}] &=& 
\hat{G}(s+\lambda)  \left[ \frac{\lambda}{\lambda-s} \hat{G}(s) - \frac{s}{\lambda-s} \hat{G}(\lambda)  \right] 					\label{aoi0}
\\ \nonumber
\mathbb{E}^0[e^{-s \alpha(S_0)} ;{K_0=1}] &=& 
\left(\hat{G}(s) - \hat{G}(s+\lambda)\right) \left[ \frac{\lambda}{\lambda-s} \hat{G}(s) - \frac{s}{\lambda-s} \hat{G}(\lambda)  \right] .		\label{aoi1}
\end{eqnarray}
Substituting the last two lines into the right hand side of (\ref{stat-palm-aoi})
we obtain
\begin{equation}  \label{Fifo-Aoi}
\mathbb{E}[e^{-s\alpha(0)}] = \frac{\hat{G}(s)  \left[ \frac{\lambda}{\lambda-s} \hat{G}(s) - \frac{s}{\lambda-s} \hat{G}(\lambda)  \right]
\left( \frac{\hat{G}(s+\lambda)}{s+\lambda} + \frac{1-\hat{G}(s)}{s} \right)}
{\frac{1}{\lambda}\left(\frac{\lambda}{\mu} + \hat{G}(\lambda)\right)},
\end{equation}
which gives  \eqref{Fifo-decomposition} if we take into account the
definition of $\hat G_I$.
\end{proof}

\begin{corollary} \label{cormeanb2}
Expression \eqref{Fifo-decomposition}
gives the stationary AoI as a sum of three independent random variables. 
In particular, the middle term in the right hand side of (\ref{Fifo-decomposition}) corresponds to the Laplace transform of the random variable
$(\sigma -\tau)^+$.
Moreover, the expectation of $\alpha(0)$ is given by
\begin{equation*} \label{mean_AoI_inside}
\mathbb{E}[\alpha(0)] \;=\; 
\frac{2}{\mu}
- \frac{1- \hat{G}(\lambda)}{\lambda} +
\frac{\hat{G}(\lambda)-\lambda \hat{G}'(\lambda) + \frac{1}{2} \lambda^2 
\int_0^\infty x^2 dG(x)}{\lambda \left(
\frac{\lambda}{\mu}
+ \hat{G}(\lambda)\right)}.
\end{equation*}
\end{corollary}
We obtained this corollary 
directly from the Laplace transform \eqref{Fifo-decomposition}
where we recognize that $\frac12 \mu \int_0^\infty x^2 dG(x)
= \int_0^\infty x dB_I(x)$. 
Notice that if the message size
has high variance then so does $\E \alpha(0)$. In particular, $\E \alpha(0) = \infty$
if $\int x^2 dG(x)=\infty$. Rather than seeing this as a problem, one should change
the point of view and adopt another function of $\alpha$ as a performance measure,
for instance, $\E \alpha(0)^p$ for some $p<1$.

\begin{corollary}
\label{corob2exp}
For  $\BB_2$, with $G$ being exponential with mean $1/\mu$ we have
\begin{equation}   \label{Fifo-Aoi-exp}
\mathbb{E}[e^{-s\alpha(0)}] \;=\; \left( \frac{\mu}{s+\mu} \right)^3 \frac{\lambda}{s+\lambda} \, \frac{s^2+2s(\lambda+\mu) + \lambda^2 + \lambda \mu +\mu^2}
{ \left( \lambda^2 + \lambda \mu + \mu^2\right)},
\end{equation}
\begin{equation} \label{meanexpb2}
\mathbb{E}[\alpha(0)] \;=\; \frac{3\lambda^3+2\lambda^2\mu+2\lambda \mu^2 + \mu^3}{\lambda \mu \left( \lambda^2 + \lambda \mu + \mu^2\right)}.
\end{equation}
\end{corollary}

Inverting the last Laplace transform gives a measure with density equal to
\eqref{fb}.
Interestingly, as  $\lambda \to \infty$ we immediately see from \eqref{Fifo-Aoi-exp} that
$\E [e^{-s\alpha(0)}] \to (\mu/(s+\mu))^3$, the Laplace transform of 
the sum of 3 i.i.d.\ exponentials. See Section \ref{sechigh} for an
explanation of this.

\section{The $\PP_2$ system}   \label{p2sec}
We remind the reader that $\PP_2$ differs from $\BB_2$ in that the arriving
message is always admitted by replacing the message (if any) sitting in the second
cell of the buffer, see Figure \ref{p2fig}.
Again, $\PP_2$ is not service-preemptive: once a message starts being
processed it will not be interrupted.
The strategy for obtaining the Laplace transform of $\alpha(0)$ is the same as before.
We make use of \eqref{stat-palm-aoi} of Proposition \ref{prop1}
which needs computation of the quantities involving $\alpha(0)$ in its right-hand side.
The analog of Lemma \ref{lemma:block} is Lemma \ref{lemma:PO}  below
which looks conspicuously the same. In fact, the first two formulas are identical.
The last two differ.

\begin{lemma} \label{lemma:PO} 
For $\PP_2$,
\begin{align} 
\label{POaoi00}
\mathbb{E}^0[e^{-s \alpha(S_0)} ;{K_{-1}=0,K_0=0}] &=  \hat{G}(\lambda)\hat{G}(s+\lambda), \\  
\label{POaoi01}
\mathbb{E}^0[e^{-s \alpha(S_0)} ;{K_{-1}=0,K_0=1}] &= \hat{G}(\lambda) \left(\hat{G}(s) - \hat{G}(s+\lambda)\right), \\    
\label{POaoi10}
\mathbb{E}^0[e^{-s \alpha(S_0)} ;{K_{-1}=1,K_0=0}] &= \frac{\lambda}{\lambda+s} \left(1- \hat{G}(s+\lambda) \right)\,\hat{G}(s+\lambda), \\  
\label{POaoi11}
\mathbb{E}^0[e^{-s \alpha(S_0)} ;{K_{-1}=1,K_0=1}] &= \frac{\lambda}{\lambda+s}\left(1-\hat{G}(\lambda+s)\right)\,  \left( \hat{G}(s)-\hat{G}(s+\lambda) \right)
\end{align}
\end{lemma}
\begin{proof}
1) When $K_{-1}=0, K_0=0$ or when $K_{-1}=0,K_0=1$ the AoI same as in the 
$\BB_2$ system, the reason being that  the number 
of messages in the system is always at most 1,
see Figure \ref{fig:lm1}. 

\no
2) Suppose next that $K_{-1}=1,K_0=0$.
In Figure \ref{fig:lm3} we depict the two scenaria corresponding
to the possible values of $K_{-2}$,
namely $(K_{-2},K_{-1},K_0)=(0,1,0)$ or $(1,1,0)$. 
In both cases, \eqref{aoidef} and the system dynamics imply that
\[
\alpha(0) = S_0-S_{-1}+V,
\]
$V$ is the time elapsed between the last arrival
in the interval $(S_{-2},S_{-1})$ and $S_{-1}$. 
If there is only one arrival in this interval then $V:=S_{-1}-T_{0}$. 
In any case, \[
\text{conditionally on $\{K_{-1}=1,K_0=0\}$, the random variables
$V$ and $S_0-S_{-1}$ are independent.}
\]
Therefore,
\[
\mathbb{E}^0[e^{-s (S_0-S_{-1}+V)} \mid K_{-1}=0,K_0=1] = \mathbb{E}^0[e^{-s (S_0-S_{-1})} \mid K_{-1}=0,K_0=1]\, \mathbb{E}^0[e^{-s V} \mid K_{-1}=0,K_0=1]
\]
and the first factor on the right is easy:
\[
\mathbb{E}^0[e^{-s (S_0-S_{-1})} \mid K_{-1}=0,K_0=1] \;=\; \int_0^\infty e^{-s x} \frac{e^{-\lambda x} dG(x)}{\hat{G}(\lambda)} \;=\; \frac{\hat{G}(s+\lambda)}{\hat{G}(\lambda)}.
\]
To evaluate $\mathbb{E}^0[e^{-s V} \mid K_{-1}=0,K_0=1]$ we note that, $V$ is the distance of the last Poisson point inside the interval $(T_{-1},S_{-1})$ (in the
left scenario of Figure \ref{fig:lm3}) or the interval $(T_{-1},S_{-1})$ in the right scenario. (In both cases the length of the interval is that of a message size
conditioned on containing at least one Poisson point.) 
To obtain the Laplace transform of $V$  look backward in time starting from $S_{-1}$  until the first Poisson point appears and condition on the event that this occurs between $S_{-1}$ and $S_{-2}$.
Thus, the density of $V$ at $v>0$ is 
\[
\frac{(1-G(v))\lambda e^{-\lambda v}}{1-\hat{G}(\lambda)},
\]
which gives
\begin{equation} \label{eqV}
\mathbb{E}^0[e^{-s V} \mid K_{-1}=0,K_0=1] \;=\; \int_0^\infty e^{-sv} \frac{\lambda e^{-\lambda v} (1-G(v))}{1-\hat{G}(\lambda)} dv \;=\; \frac{\lambda}{\lambda+s} \, \frac{1-\hat{G}(s+\lambda)}{1-\hat{G}(\lambda)} .
\end{equation}
Putting these together we obtain \eqref{POaoi10}.

\begin{figure}[h]
  \begin{center}
    \includegraphics[width=4.5in]{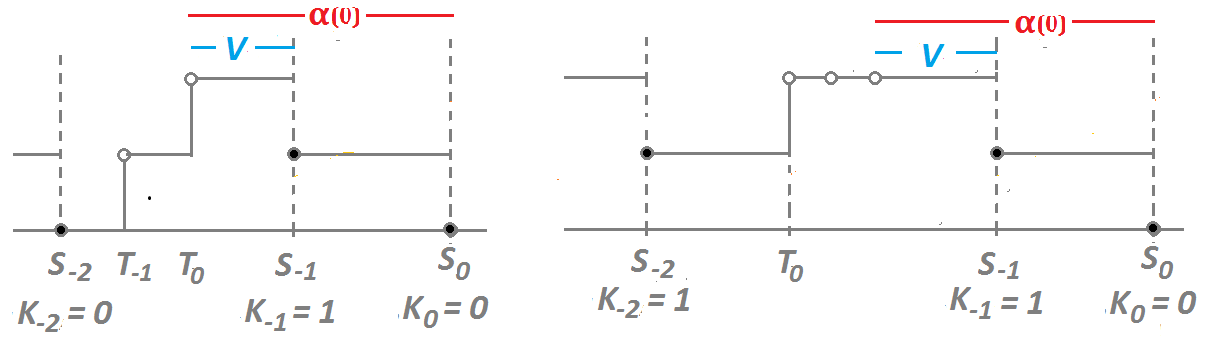}
  \end{center}
  \caption{ Under $\PP_2$, the segments $[S_{-2},S_{-1})$, $[S_{-1},S_0)$  when $K_{-1}=1,K_0=0$ in the case where $K_{-2}=0$ (left) and $K_{-2}=1$ (right).}
  \label{fig:lm3}
\end{figure}

\no
3) Finally, assume that $K_{-1}=1,K_0=1$.
This situation is similar to the previous one and thus will be treated succinctly. 
We are guided by Figure \ref{fig:lm4}.  Firstly, we have
\[
\mathbb{E}^0[e^{-s (S_0-S_{-1})} \mid K_{-1}=1,K_0=1] \;=\; \int_0^\infty e^{-s x} \frac{1-e^{-\lambda x} dG(x)}{1-\hat{G}(\lambda)} \;=\; \frac{\hat{G}(s)-\hat{G}(s+\lambda)}{1-\hat{G}(\lambda)}.
\]
\begin{center}
\begin{figure}[h]
  \begin{center}
    \includegraphics[width=4.5in]{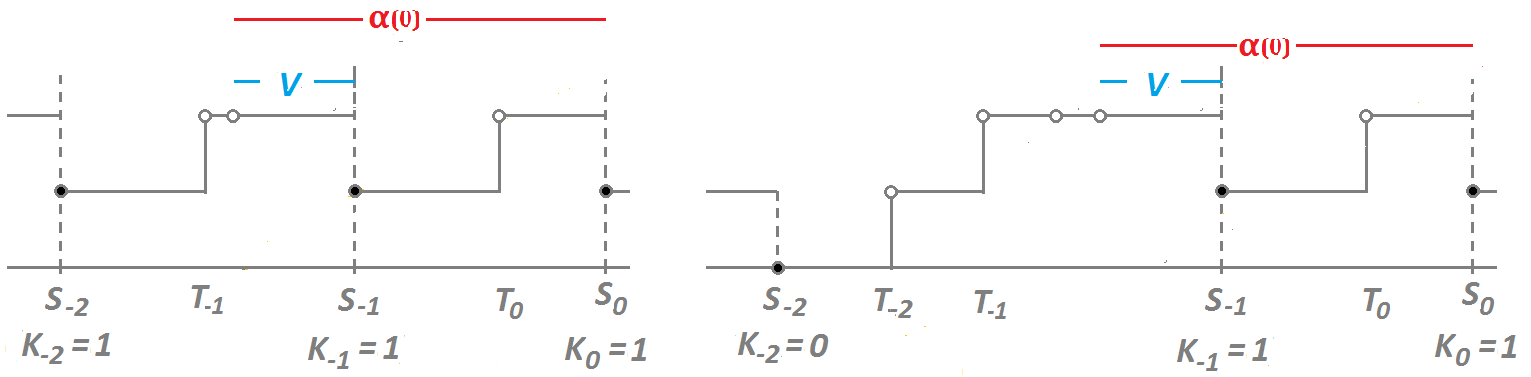}
  \end{center}
  \caption{ Under $\PP_2$, the segments $[S_{-2},S_{-1})$, $[S_{-1},S_0)$  when $K_{-1}=1,K_0=1$ in the case where $K_{-2}=0$ (left) and $K_{-2}=1$ (right).}
  \label{fig:lm4}
\end{figure}
\end{center}
Secondly, the argument used to derive (\ref{eqV}) can be used here 
again with no changes to obtain
\[
\mathbb{E}^0[e^{-s V} \mid K_{-1}=1,K_0=1] \;=\; \int_0^\infty e^{-sv} \frac{\lambda e^{-\lambda v} (1-G(v))}{1-\hat{G}(\lambda)} dv
\;=\; \frac{\lambda}{\lambda+s} \, \frac{1-\hat{G}(s+\lambda)}{1-\hat{G}(\lambda)} .
\]
Putting these together we obtain \eqref{POaoi10} as well.
\end{proof}

The formula for $\E e^{-s \alpha(0)}$ now clear.
\begin{theorem}\label{thmP2} 
For $\PP_2$,
the Laplace transform of the stationary Age of Information is given by
\begin{equation}  \label{PO_Aoi_distr-2}
\mathbb{E}[e^{-s \alpha(0)}]  \;=\;  \hat{G}(s) \, \left( \hat{G}(\lambda) + \frac{\lambda}{\lambda+s} \left(1-\hat{G}(s+\lambda)\right)  \right) \,
\left( \frac{ \hat{G}(\lambda)}{\frac{\lambda}{\mu} +  \hat{G}(\lambda)}  \frac{\lambda}{\lambda+s}
\frac{\hat{G}(s+\lambda)}{\hat{G}(\lambda)} + \frac{\frac{\lambda}{\mu}}{\frac{\lambda}{\mu} +  \hat{G}(\lambda)} \hat{G}_I(s) \right)
\end{equation}
\end{theorem}
\begin{proof}
Adding up (\ref{POaoi00}) and (\ref{POaoi01}) of Lemma \ref{lemma:PO} and similarly (\ref{POaoi10}) and (\ref{POaoi11}) we obtain
\begin{align*}
\mathbb{E}[e^{-s \alpha(S_0)} ;{K_0=0}] 
&= \hat{G}(s+\lambda) \left[ \hat{G}(\lambda) + \frac{\lambda}{\lambda+s} \left(1- \hat{G}(s+\lambda) \right) \right] 					
\\			
\mathbb{E}[e^{-s \alpha(S_0)}   ;{K_0=1}   ] 
&=  \left( \hat{G}(s)-\hat{G}(s+\lambda) \right)  \, \left[ \hat{G}(\lambda)  + \frac{\lambda}{\lambda+s}\left(1-\hat{G}(\lambda+s)\right)\right].
\end{align*}
Substituting these expressions  in  the numerator of (\ref{stat-palm-aoi}), and recalling the definition \eqref{GI} of $G_I$, 
we obtain \eqref{PO_Aoi_distr-2}.
\end{proof}
An alternative expression for \eqref{PO_Aoi_distr-2} is:
\begin{eqnarray}
\mathbb{E}[e^{-s\alpha(0)}] &=& \frac{\hat{G}(s) \left( \hat{G}(\lambda) + \frac{\lambda}{\mu} \hat{G}_I(s+\lambda) \right) \left( \frac{\lambda}{\lambda+s}\hat{G}(s+\lambda) + \frac{\lambda}{\mu} \hat{G}_I(s)\right)}
{\frac{\lambda}{\mu}+\hat{G}(\lambda)} .\label{PO_Aoi_distr}
\end{eqnarray}

\begin{corollary} \label{cormeanp2}
In expression (\ref{PO_Aoi_distr-2}) we recognize that $\alpha(0)$
is equal in distribution to the sum of three independent random variables,
of which the middle one,
$\hat{G}(\lambda) + \frac{\lambda}{\lambda+s} \left(1-\hat{G}(s+\lambda)\right)$, 
is the Laplace transform of the random variable $\tau \1_{\tau>\sigma}$.
Moreover, 
\begin{equation}  \label{PO_Aoi_mean}
\mathbb{E}[\alpha(0)] \;=\; \frac{1}{\mu}+ \frac{1}{\lambda}\left( 1-\hat{G}(\lambda) + \lambda \hat{G}'(\lambda) \right) + \frac{1}{\lambda}\, \frac{1}{\frac{\lambda}{\mu}+\hat{G}(\lambda)} \left( \hat{G}(\lambda) - \lambda \hat{G}'(\lambda)
+ \frac{1}{2} \lambda^2 \hat{G}''(0) \right)
\end{equation}
\end{corollary}
One should compare this to the expectation for the $\PP_1$ system, \cite{Kosta17,KKZ19,Inoue19},
\[
\mathbb{E}[\alpha(0)] \;=\; \frac{1}{\lambda \hat{G}(\lambda)}.
\]
\begin{corollary}
\label{corop2exp}
For $\PP_2$ with exponential message sizes,
\begin{equation}  			\label{MM-AoI-distr}
\mathbb{E}[e^{-s\alpha(0)}] 
=  \frac{\mu}{\mu+s} \, 
\left( \frac{\mu}{\mu+\lambda} + \frac{\lambda}{\lambda+\mu+s} \right) \,
\left( \frac{\mu^2}{\lambda^2+\lambda \mu + \mu^2} \frac{\lambda}{\lambda+s} \frac{\lambda+\mu}{\lambda+\mu+s} + \frac{\lambda^2+\lambda \mu}{\lambda^2+\lambda \mu + \mu^2} \frac{\mu}{\mu+s} \right)   ,
\end{equation}
\begin{equation}  	  \label{meanexpp2}
\mathbb{E}[\alpha(0)] 
= \frac{2\lambda^{5}+7\lambda^{4}\mu +8\lambda^{3}\mu^{2}+7\lambda^{2}\mu^{3}+4\lambda \mu^{4}+\mu^{5}}{\lambda \mu \left( \lambda +\mu \right)^{2}\left( \lambda^{2}+\lambda \mu +\mu^{2}\right) }		,
\end{equation}
and, with $\rho=\lambda/\mu$, the standard deviation of $\alpha(0)$ under $\P$ is
\[
\operatorname{sd}_\P (\alpha(0)) 
= \frac 1 \mu  \frac{\sqrt{2\rho ^{10}+12\rho ^{9} +35\rho ^{8} +60\rho
^{7} +66\rho ^{6} +56\rho ^{5} +45\rho ^{4}
+34\rho ^{3} +18\rho ^{2} +6\rho  + 1}}
{\rho  \left( \rho +1 \right) ^2   \left( \rho
^{2}+\rho  +1\right) }.
\]
\end{corollary}

The expectation 
(consistent with \cite{Kosta17} in this case)
and variance have been computed by summing up the expectations
and variance of the three independent random variables comprising $\alpha(0)$.
Inverting $\E e^{-s \alpha(0)}$ shows that $\alpha(0)$ has the density \eqref{fp}.
It is easy to see from \eqref{MM-AoI-distr} that
$\lim_{\lambda} \to \infty \E[e^{-s\alpha(0)}] = (\mu/(s+\mu))^2$,
the sum of 2 i.i.d.\ exponentials. See Section \ref{sechigh}.

\section{Comparisons and optimality}
\label{compa}
Slightly abusing notation, we write $\alpha_{\PP_n}$ instead of $\alpha_{\PP_n}(0)$;
To further simplify life, we shall now use normalized units, assuming $\mu=1$.
We draw conclusions from the work above, as well as the results of \cite{Kosta17,Inoue19,KKZ19},
attempting to justify our claim that ``small buffers suffice''.
We first summarize observations regarding $\PP_n, \BB_n$, $n=1,2$,
and then consider larger $n$.

\subsection{Recalling formulas for $\PP_1$ and $\BB_1$}
Concerning the $\PP_1$ system we have, from \cite[Corollary 4]{KKZ19},
\begin{equation} \label{p1}
\E[ e^{-s\alpha_{\PP_1}} ]=  \frac{\rho \hat G(s+\rho)}{s+\rho \hat G(s+\rho)},
\quad 
\E[ \alpha_{\PP_1} ] = \frac{1}{\rho \hat G(\rho)}.
\end{equation}
On the other hand, for $\BB_1$, \cite[Corollary 9(i)]{KKZ19}, gives
\begin{equation} \label{b1}
\E[ e^{-s\alpha_{\BB_1}} ]= \frac{\rho}{1+ \rho} \cdot
\frac{(s+\rho-\rho \hat G (s)) \hat G  (s)}{s(s+\rho)},
\quad
\E [\alpha_{\BB_1}] 
=1+ \frac{1}{\rho} + \frac{\rho}{2}\cdot
\frac{\E \sigma^2}{1+\rho}.
\end{equation}

We now have information about all systems that we now compare.
The comparisons depend on the message size distributions. We
choose to consider two ``extremes''. First, exponentially distributed size,
second, deterministic, representing maximal and minimal randomness.
The observations are summarized in plots
rather than formulas because the latter, albeit explicit in almost all cases 
(but see \S\ref{ap1} below for an exception), are not succinctly presentable.

\subsection{Exponential message sizes}
We obtain explicit formulas from Corollary \ref{corob2exp} for $\alpha_{\BB_2}$,
Corollary \ref{corop2exp} for $\alpha_{\PP_2}$ and \eqref{b1}, \eqref{p1},
for $\alpha_{\BB_1}$, $\alpha_{\PP_1}$, respectively.
Using the notation $M_{\PP_1}(\rho)$ for $\E [\alpha_{\PP_1}]$, 
where $\rho=\lambda/\mu=\lambda$, in normalized units.
Similarly for other systems.
We summarize the comparisons in a plot:
\begin{center}
\includegraphics[width=0.4\textwidth]{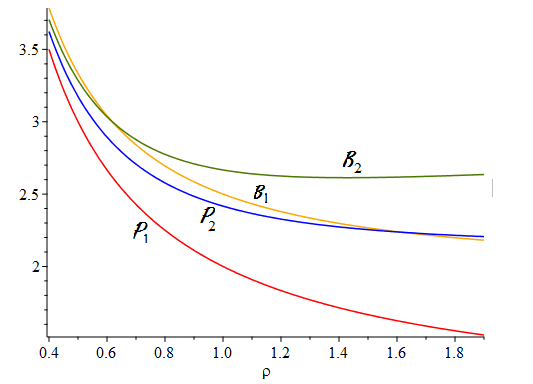}
\includegraphics[width=0.4\textwidth]{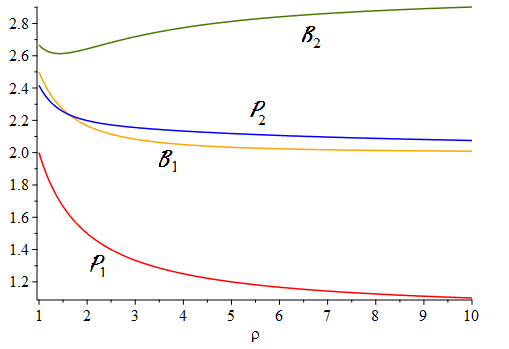}
\captionof{figure}{Mean AoI as a function of $\rho$;
the left plot is for $\rho < 1$;
the right is for $\rho \ge 1$.} 
\label{meansexp}
\end{center}
We see that
\[
M_{\PP_1}(\rho) < M_{\PP_2}(\rho) < M_{\BB_2}(\rho) \text{ for all $\rho$. }
\]
The odd system is $\BB_1$.
For small $\rho$, $M_{\BB_1}(\rho)$ is worst (highest).
For large $\rho$, $M_{\BB_1}(\rho)$ is between 
$M_{\PP_1}(\rho)$ and $M_{\PP_2}(\rho)$.
There is also an intermediate zone, where $M_{\BB_1}(\rho)$ is between 
$M_{\PP_2}(\rho)$ and $M_{\BB_2}(\rho)$.

We can also ask whether the comparisons above remain true
in the sense of stochastic ordering. Recall that a real random variable
$X$ is stochastically smaller than $Y$,
and write 
\[
\text{$X \sto Y$, if $\P(X>u) \le \P(Y>u)$ for all $u \in \R$.}
\]
Note that stochastic ordering is a partial order in the space of probability measures
on the real line so two random variables may not be comparable at all.
The information is obtained by inverting the Laplace transforms
\eqref{Fifo-Aoi-exp} and \eqref{MM-AoI-distr} which give the densities
\eqref{fb} and \eqref{fp}, respectively. It is also easy to invert the Laplace transforms
\eqref{p1} and \eqref{b1}. Integrating the densities from $t$ to $\infty$,
we obtain the complementary distribution functions, better summarized  
in a couple of plots:
\begin{center}
\includegraphics[width=0.4\textwidth]{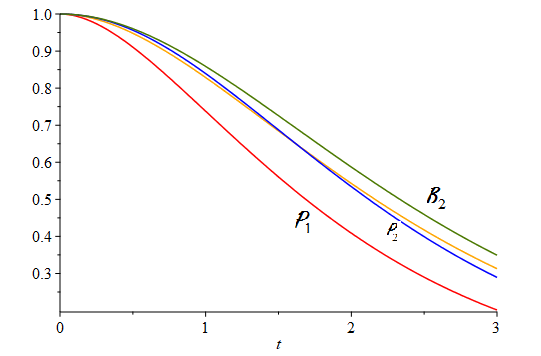}
\includegraphics[width=0.4\textwidth]{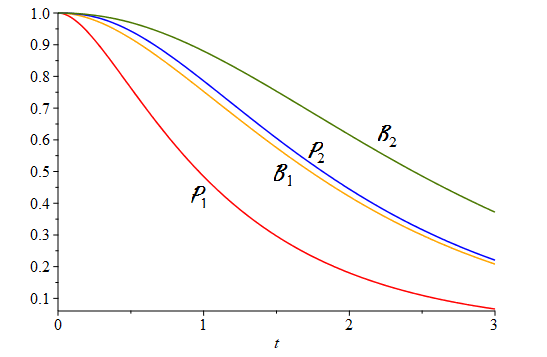}
\captionof{figure}{ $\P(\alpha>t)$ plotted against $t$ for
small $\rho$ on the left and high $\rho$ on the right.} 
\label{tailsexp}
\end{center}
We obtain that
\[
\alpha_{\PP_1} \sto \alpha_{\PP_2} \sto \alpha_{\BB_2} \text{ for all } \rho.
\]
Moreover,
\[
\alpha_{\PP_1} \sto \alpha_{\BB_1} \sto \alpha_{\PP_2}
\text{ for all sufficiently high } \rho.
\]
The following figure gives plots of variances as functions of $\rho$.
\begin{center}
\includegraphics[width=0.4\textwidth,height=4cm]{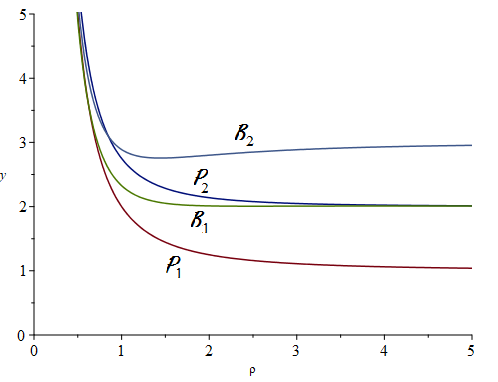}
\captionof{figure}{Variances as a functions of $\rho$}
\label{varexp}
\end{center}
Note that they all converge to integers.

\subsection{Deterministic message sizes}
\label{detsizes}
We now assume that $\P(\sigma=1)=1$: message sizes are all equal to $1$
with probability $1$. We can thus easily obtain $M(\rho)$ in all cases
by setting $\sigma=1$ in the formulas of Corollaries \ref{cormeanb2} and
\ref{cormeanp2} and in \eqref{b1} and \eqref{p1}.
They are summarized in Figure \ref{meansdet}.
We observe that
\[
M_{\PP_2}(\rho) < M_{\BB_2}(\rho) < M_{\PP_1}(\rho)
\text{ for all $\rho$. }
\]
\begin{center}
\includegraphics[width=0.4\textwidth]{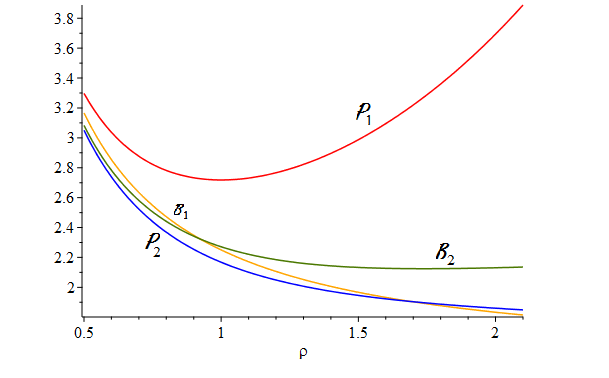}
\includegraphics[width=0.4\textwidth]{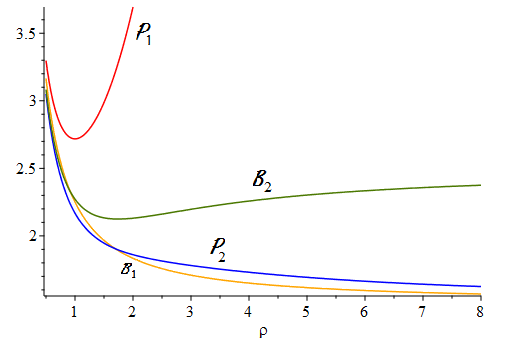}
\captionof{figure}{ Mean AoI as a function of $\rho$;
right plot extends to high values of $\rho$} 
\label{meansdet}
\end{center}
Whereas $\PP_1$ was best in the exponential case, it is now worst.
In fact, as 
\[
\lim_{\rho \to \infty} M_{\PP_1}(\rho) = \infty.
\]
The worst system, from the point of view of expectation,
is thus $\PP_1$. However, as in the exponential case,
$\BB_1$ is the odd system in that it is between $\BB_2$ and $\PP_1$
for small $\rho$, but $M_{\BB_1}(\rho) < M_{\PP_2}(\rho)$ for all
large enough $\rho$.
However, the difference between the two goes to $0$ as $\rho \to \infty$.
We can easily see that $\lim_{\rho\to\infty} M_{\BB_1}(\rho)
= \lim_{\rho\to\infty} M_{\PP_2}(\rho) = 3/2$, while
$\lim_{\rho\to\infty} M_{\BB_2}(\rho) = 5/2$.

We again ask whether the comparisons in the mean translate to stochastic
comparisons. 
We observe that
\[
\alpha_{\PP_1} \xrightarrow[\rho \to \infty]{\text{\rm d}} \infty.
\]
The reason for this is clear: 
when $\rho$ is high, the message being processed is constantly
interrupted. Since the message size is always $1$ no message has a chance to
ever be completed.
To obtain information about $\P(\alpha_{\PP_1}> x)$ for all $x$, we resort to
numerics as the Laplace transform \eqref{p1} with 
$\hat G(s) = e^{-s}$ is not invertible. For further discussion on
the distribution of $\alpha_{\PP_1}$ see \S \ref{ap1} below.
Luckily, the Laplace transforms for all other variables, $\alpha_{\BB_1},
\alpha_{\BB_2}, \alpha_{\PP_2}$ are all invertible and  
correspond to random variables with densities
that can all be analytically computed.
We summarize the comparisons of the distributions in the plot below.
\begin{center}
\includegraphics[width=0.4\textwidth]{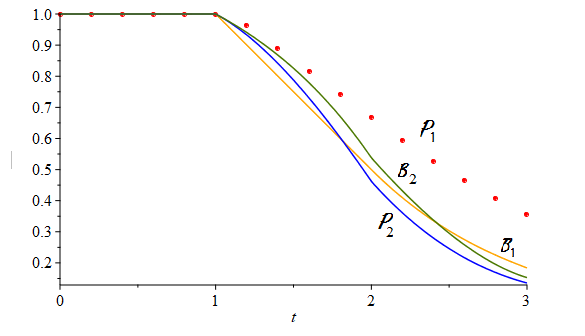}
\includegraphics[width=0.4\textwidth]{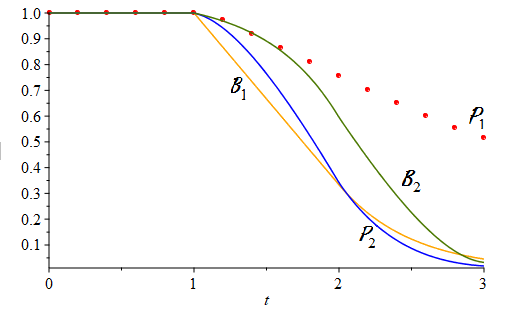}
\captionof{figure}{ $\P(\alpha>t)$ plotted against $t$ for
small $\rho$ on the left and high $\rho$ on the right.} 
\label{tailsdet}
\end{center}
Our observation is then that
\[
\alpha_{\PP_2} \sto \alpha_{\BB_2} \sto \alpha_{\PP_1} , \,\,
\alpha_{\BB_1} \sto \alpha_{\PP_1} \text{ for all } \rho,
\]
whereas $\alpha_{\BB_1}$ is not comparable to
any of the other three random variables.
Figure \ref{densdet} shows the densities for $\alpha_{\PP_2}$ and $\alpha_{\BB_2}$ 
for various values of $\rho$.
\begin{center}
\includegraphics[width=0.4\textwidth,height=3cm]{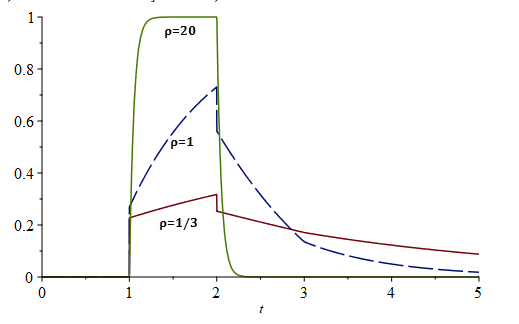}
\includegraphics[width=0.4\textwidth,height=3cm]{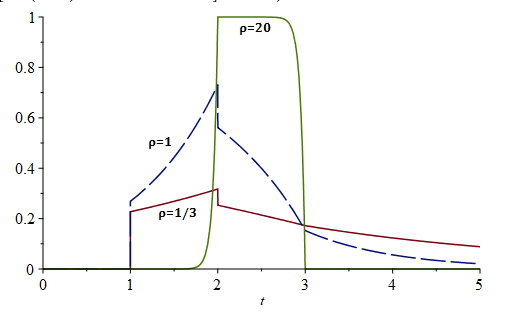}
\captionof{figure}{Densities of $\alpha_{\PP_2}$ (left) 
and $\alpha_{\BB_2}$ (right)  for various traffic intensities.} 
\label{densdet}
\end{center}
Variance plots are in Figure \ref{vardet}.
\begin{center}
\includegraphics[width=0.4\textwidth,height=3cm]{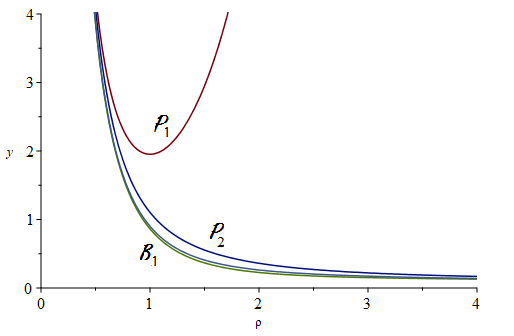}
\captionof{figure}{variances all systems as a function of $\rho$}
\label{vardet}
\end{center}
We have
$\lim_{\rho\to\infty} \var_\P (\alpha_{\BB_1}) =  \lim_{\rho\to\infty} \var_\P (\alpha_{\PP_2}) = \lim_{\rho\to\infty} \var_\P (\alpha_{\BB_2})
= 1/12$.

\subsection{High traffic asymptotics}  \label{sechigh}
``High traffic asymptotics" refers to the regime $\rho \to \infty$.
Even though we have no explicit formulas for $\PP_n$ or $\BB_n$ 
when $n \ge 3$, we can easily obtain asymptotics from the system dynamics.
\begin{proposition} \label{proplim}
Let $\sigma, \sigma_1, \sigma_2, \ldots$ be i.i.d.\ copies of $\sigma$.
Let $\sigma_I$ be distributed as $\hat G_I$ as in \eqref{GI}.
Then
\[
\alpha_{\PP_n} \xrightarrow[\rho \to \infty]{\text{\rm d}} \sigma + \sigma_I, \quad 
n \ge 2,
\]
while
\[
\alpha_{\BB_n} \xrightarrow[\rho \to \infty]{\text{\rm d}} \sigma_1+\cdots + \sigma_n 
+ \sigma_I, \quad 
n \ge 1.
\]
\end{proposition}
\begin{proof}[Sketch of proof]
In both systems, the buffer consists of $n$ cells. The message being processed
sits in cell $1$. In $\PP_n$, the freshest message is either in cell 1, in which case
all other cells are empty, or in cell 2. When $\rho$ is high there is always a
message being processed in cell 1 and the freshest message is in cell 2.
Hence, at any time $t$, the AoI $\alpha_{\PP_n}(t)$ equals the 
service time of the message in cell 2 plus the remaining 
service time of the message in cell 1.
These are two independent random variables. The first one is distributed
as $\sigma$. The second is distributed as $\sigma_I$ since the system is
stationary.
For $\BB_1$, we can obtain the limit from the Laplace transform of 
\eqref{b1}. It is easy to to see that $\lim_{\rho \to \infty}
\E [e^{-s \alpha_{\BB_1}}] = \hat G(s) \, \hat G_I(s)$
and so $\alpha_{\BB_1} \xrightarrow[\rho \to \infty]{\text{\rm d}} \sigma
+ \sigma_I$.
For general $n$, when $\rho$ is high, the AoI $\alpha_{\BB_n}(t)$
equals the remaining service time of the message in cell 1 (in distribution equal to
$\sigma_I$) plus the time
elapsed until the beginning of its service which is, in distribution,
equal to the sum of $n$ independent service times. 
\end{proof}

\begin{remark}\rm
When $\sigma=1$ with probability $1$, $\sigma_I$ is a  uniform random variable
in the interval $[0,1]$. Hence $\sigma_1+\cdots+\sigma_n+\sigma_I
= n + \sigma_I$ and the variance of this random variable is $1/12$,
in agreement with the observation around Figure \ref{vardet} for $\BB_1$
and $\BB_2$.
Similarly, for $\PP_2$, the asymptotic variance is again $1/12$.
In fact, the limits in Proposition \ref{proplim} also explain the asymptotic shapes
of the densities in Figure \ref{densdet}.
\\
The limits of the variances in Figure \ref{vardet} for $\BB_1, \BB_2, \PP_2$ are also 
explainable in the light of the proposition above.
\end{remark}

It remains to see what the limit of $\alpha_{\PP_1}$ is as $\rho \to \infty$.
This should be considered separately as the limit very much depends on
the distribution of $\sigma$. 
If $\sigma=1$ with probability 1, we explained above why $\alpha_{\PP_1}$
converges to a random variable that takes value $\infty$ with probability 1.
In fact, this should be true if $\sigma$ has a distribution whose support is at positive 
distance from $0$.
If $\sigma$ is exponential with mean 1, we see directly from \eqref{p1},
that $\lim_{\rho \to \infty} \E[e^{-s\alpha_{\PP_1}}] = 1/(s+1)$.
Thus the limit of $\alpha_{\PP_1}$ depends both on the tail
of the distribution and on its behavior at $0$.

\subsection{On the distribution of $\alpha_{\PP_1}$ for deterministic message size}
\label{ap1}
We now complete the discussion in \S\ref{detsizes} by giving some more information
on $\alpha_{\PP_1}$, when $\sigma=1$ with probability 1, 
whose distribution is not explicitly computable but
has some interesting properties, including a combinatorial explanation of 
all of its moments.
Another reason we devote a little more ink on $\alpha_{\PP_1}$ is that it seems to be an
upper bound on the age of information so long as message sizes
are close to being deterministic.
From \eqref{p1} with $\hat G(s) = \exp(-s)$, we have
\[
\E[e^{-s \alpha_{\PP1}}] = \frac{1}{m\, s e^s +1} =:  L_m(s),
\]
with 
\[
m:= e^\rho/\rho= \E[\alpha_{\PP_1}].
\]
This Laplace transform cannot be analytically inverted.
However, since $L_m(s)$ as a function of a complex variable $s$
has no singularity on the closed right-half plane $\Re(s) \ge 0$,
we used the inversion formula \cite[\S 8.2]{MH}
\[
f_m(t) = \frac{1}{2\pi} \int_{-\infty}^\infty \frac{e^{iy}}{1+im y e^{iy}} dy,
\]
together with a numerical approximation of this extended Riemann integral
in order to obtain information about the tail of the distribution of
$\alpha_{\PP_1}$ which is plotted in Figure \ref{tailsdet} as a dotted line.
We can make a few remarks about $f_m$.

First, it is obviously a probability density function for all $m \ge e$ since we know
that $\inf_{\rho>0} e^\rho/\rho = e$.
One might conjecture that $f_m$ remains a probability density function for
all $m \ge 0$. 
However, $L_m$ seizes to be completely monotone 
\cite[XIII.4]{FEL}
for $m$ small enough
and thus, the conjecture is false. 
To see this, it is easy to see that
\[
L''_m(s) < 0 \text{ for $s$ in a neighborhood
of zero if (and only if) } m < 1.
\]
We strengthen the conjecture by claiming that
\begin{conjecture}
Let $m\ge 0$.
Then $L_m(s) = 1/(m s e^s+1)$ 
is the Laplace transform of some random variable if and only if $m=0$
(corresponding to a trivial random variable) or $m \ge e$.
\end{conjecture}

Second, since $L_m(s)$ is infinitely differentiable at all $s \ge 0$,
we have that all moments of $\alpha_{\PP_1}$  exist.
Using a formal series expansion, we can easily express the $p$-th moment in terms of
the polynomial
\[
Q_p(z) := \sum_{k=1}^p (p)_k k^{p-k} z^k, \quad p=1,2,\ldots
\]
as follows
\[
\E [\alpha_{\PP_1}^p] = (-1)^p Q_p(-m),
\]
with $m= e^\rho/\rho$, and then show that this is correct.
It is easy to see that $Q_p(z)$ has a combinatorial interpretation.
It is the generating function of the sequence $(p)_k\, k^{p-k}$,
$1 \le k \le p$, where $(p)_k = p(p-1)\cdots (p-k+1)$, counting the
number of ways to form $k$ labeled groups, 
each with a distinct leader, using $p$ different people.
For values and other properties, see \cite{OEIS}.
One can also see that, for $p$ large, 
\[
\E [\alpha_{\PP_1}^p] \approx p! m^p,
\]
that is, the
$p$-th moment of an exponential 
random variable with the same mean.

\subsection{The best systems} \label{secbest}

We now return to the problem of choosing the best system so that we keep
AoI ``as small as possible''.
Even though we have no proof for the optimal system, we have enough reasons
to justify making the following conjecture.
\begin{conjecture}
Consider the collection $\PP_n, \BB_n$, $n =1,2,\ldots$
Assume that each system is driven by a Poisson arrival process of rate $\lambda$.
Then, regardless of the message size distribution,
\[
\alpha_\II  \sto \alpha_\JJ 
\]
for all  $\II \in \{\PP_1, \BB_1, \PP_2, \BB_2\}$
and all $\JJ \in \{\PP_n, \BB_n, \, n \ge 3\}$.
\end{conjecture}
Evidence for this conjecture is provided by the results of \cite{Inoue19,KKZ19},
the results in this paper and the following observations.
First, it is clear that
\[
\alpha_{\PP_n}(t) \le \alpha_{\BB_n}(t), \text{ for all } t \text{ and all } n \ge 2
\]
provided that all systems are driven by identical arrival processes and the
same sequence of message sizes. In other words, this is a pathwise inequality.
Similarly,
\[
\alpha_{\BB_2}(t) \le \alpha_{\BB_3}(t) \le \cdots \le \alpha_{\BB_\infty}(t), \text{ for all } t.
\]
One can show that we cannot expect a similar pathwise inequality 
for the $\PP_n$ systems. 
But when $\rho$ is sufficiently large, we showed that,
for all $n \ge 2$, $\alpha_{\PP_n}(t)$ is approximately distributed as
$\sigma+\sigma_I$, so there is no reason to believe that $\alpha_{\PP_{n+1}}$
will improve $\alpha_{\PP_n}$ at high $\rho$.
A crucial step in proving the veracity of the last conjecture
would be to show that, in stationarity,
\[
\alpha_{\PP_2} \sto \alpha_{\PP_3} \sto \cdots 
\]
Thus if the conjecture is true then
it would be irrelevant to consider
any system other than
$\PP_1, \PP_2, \BB_1,\BB_2$
 insofar as stochastically minimizing the stationary AoI is concerned. 
However, see \cite{dynamic}.

\section{Some final words}\label{sec:summary}

Using the Markov embedding and Palm inversion formula,
we derived the stationary AoI distribution of $\BB_2$
and $\PP_2$ under Poisson message arrivals and generally
distributed message sizes (processing times).
The AoI of these systems and two other systems
with low AoI, $\BB_1$ and $\PP_1$, were compared  for
exponential and deterministic service times, which, in some sense,
are two extremes.
We paid particular importance to their complementary distribution functions which
play a role in important performance requirements.
$\PP_1$ has generally lower AoI for exponentially distributed message
sizes,
while $\PP_2$ had lower AoI for deterministic
message sizes when the traffic load was sufficiently low.
The performance of $\BB_1$ is interesting, having smallest AoI among these
four considered policies in some cases.
The $\BB_2$ system may need to be used instead of these other
three owing to technological constraints.

The rules of thumb derived can roughly be summarized as follows:
If the message sizes are deterministic or nearly so then it is best to
use $\BB_1$ or $\PP_2$, which is pretty close to $\BB_1$.
In particular, $\PP_1$ seems to be an upper bound for AoI
and this is the reason that we paid some special attention to it in 
\S \ref{ap1} (where we had to resort to analytical tricks to approximate it,
as there is no closed-form formula for the distribution).
On the other hand, if message sizes are ``very random'',
we expect the opposite: $\PP_1$ performs best stochastically so.
Granted, the study in this paper has been done only for Poisson arrivals and so,
to be able to analyze general renewal arrivals and service times
 one should use a different method
which we leave for future work. We should also mention that dynamic policies
should also be studied, that is, policies that decide on rejection or not 
of a message based on information such as the time the message has spent in the system,
the arrival process up to the current time, etc.  \cite{dynamic}.


Finally, we also mention that the AoI $\alpha$ defined in \eqref{aoidef}
may not be the most appropriate measure of freshness as it incorporates information
about the arrival process as well. A different measure \cite{KKZ19}
is $\beta(t) = A(t) - A(D(t))$,
in the notation of the processes introduced in \eqref{poi1} and \eqref{poi2},
the distribution of which may differ significantly from that of $\alpha$.
Also, it may be interesting to
study other  performance criteria for AoI-sensitive applications
including Cost of Update Delay \cite{Kosta17,Inoue19} or
message blocking probability.

\bibliographystyle{plain}

\vspace*{1cm}

\begin{center}
\begin{tabular}{ccc}
\begin{minipage}[t]{0.3\textwidth}
{\sc George Kesidis}\\ 
\small
Computer Science Dep, 
The Pennsylvania State University, University Park, PA 16802, USA;
\href{mailto:gik2@psu.edu}{gik2@psu.edu}
\end{minipage}
&
\begin{minipage}[t]{0.3\textwidth}
{\sc Takis Konstantopoulos}\\
\small
Department of Mathematical Sciences, 
The University of Liverpool, Liverpool  L69 7ZL, UK;
\href{mailto:takiskonst@gmail.com}{takiskonst@gmail.com}
\end{minipage}
&
\begin{minipage}[t]{0.33\textwidth}
{\sc Michael A.\ Zazanis}\\ 
\small
Department of Statistics, Athens University of Economics and Business, 76 Patission St., Athens 104 34, Greece;
\href{mailto:zazanis@aueb.gr}{zazanis@aueb.gr}
\end{minipage}
\end{tabular}
\end{center}

\end{document}